# Vacancy defects induced changes in the electronic and optical properties of NiO studied by spectroscopic ellipsometry and first-principles calculations

Kingsley O. Egbo,[1] Chao Ping Liu,[1,2] Chinedu E. Ekuma,[3,4] Kin Man Yu[1] [*]

[1]Department of Physics, City University of Hong Kong, 83 Tat Chee Avenue, Kowloon, Hong Kong
[2] Research Center for Advanced Optics and Photoelectronics, Department of Physics, College of Science, Shantou University, Shantou, Guangdong 515063, China
[3]Department of Physics, Lehigh University, Bethlehem, Pennsylvania 18015, United States
[4]Institute for Functional Materials and Devices, Lehigh University, Bethlehem, Pennsylvania, 18015 United States

## ABSTRACT

Native defects in semiconductors play an important role in their optoelectronic properties. Nickel oxide (NiO) is one of the few wide-gap *p*-type oxide semiconductors and its conductivity is believed to be controlled primarily by Ni-vacancy acceptors. Herein, we present a systematic study comparing the optoelectronic properties of stoichiometric NiO, oxygen-rich NiO with Ni vacancies (NiO:$V_{Ni}$) and Ni-rich NiO with O vacancies (NiO:$V_O$). The optical properties were obtained by spectroscopic ellipsometry while valence band spectra were probed by high-resolution X-ray photoelectron spectroscopy. The experimental results are directly compared to first-principles DFT + *U* calculations. Computational results confirm that gap states are present in both NiO systems with vacancies. Gap states in NiO:$V_O$ are predominantly Ni 3*d* states while those in NiO:$V_{Ni}$ are composed of both the Ni 3*d* and O 2*p* states. The absorption spectra for the NiO:$V_{Ni}$ sample show significant defect-induced features below 3.0 eV compared to NiO and NiO:$V_O$ samples. The increase in sub-gap absorptions in the NiO:$V_{Ni}$ can be attributed to gap states observed in the electronic density of states. The relation between native vacancy defects and electronic and optical properties of NiO are demonstrated, showing that at similar vacancy concentration, the optical constants of NiO:$V_{Ni}$ deviate significantly from those of NiO:$V_O$. Our experimental and computational results reveal that although $V_{Ni}$ are an effective acceptors in NiO, they also degrade the visible transparency of the material. Hence, for transparent optoelectronic device applications, an optimization of native $V_{Ni}$ defects with extrinsic doping is required to simultaneously enhance *p*-type conductivity and transparency.

*kinmanyu@cityu.edu.hk*

## INTRODUCTION

Nickel oxide (NiO) is a material with vast technological potential and has been widely studied as a prototypical *p*-type transition metal (TM) oxide[1,2]. It is an antiferromagnetic oxide with its Neel temperature, $T_N$ at about 523K. Previously described as a Mott-Hubbard insulator, NiO is now best characterized as a strongly correlated charge-transfer insulator, with its O 2*p* band lying between the Ni 3*d* upper and lower Hubbard bands[3]. The exciting nature of the electronic structure of NiO, makes it an extensively studied system[4–8]. More recently, NiO has also found many applications in optoelectronic devices. For example, *p*-type NiO has been broadly applied as the hole transport layer in solar cells[9–12], ultraviolet photodetectors[13–15], transparent junction diodes[16], thin-film transistors[17] and visible light transparent solar cells[18,19]. On the other hand, semi-insulating stoichiometric NiO and Ni-rich NiO have been explored as major component materials for resistive switching and capacitance modulation applications[20–22]. These technological potentials have continued to attract research interests in NiO, with particular efforts to better understand its doping[23] and transport properties[24].

Native defects play a central role in controlling the electrical and optical properties of oxide semiconductors[25,26], for example, Shook *et. al.,* recently showed the effects of vacancies on the optoelectronic properties of Cu and Ag-based p-type transparent conducting oxides[27]. In NiO, optoelectronic and transport properties are strongly affected by its native defects[28]. It has been demonstrated that Ni vacancy ($V_{Ni}$) defects lead to *p*-type conductivity in NiO. Lany *et al.*[29] and Dawson *et al.*[20] showed that under an O-rich growth condition, the defect formation energy of $V_{Ni}$ is very low while corresponding hole killers such as oxygen vacancies $V_O$ have high formation energy. These findings are consistent with the observed *p*-type conductivity observed in O-rich NiO materials[30,31]. Moreover, native vacancy defects have also been shown to strongly affect the optical character and bandgap nature of NiO. Ni-vacancy ($V_{Ni}$) states were believed to cause a degradation in its visible light transparency. Newman and Chrenko[32] noted that NiO with excess oxygen appeared darker and they attributed this to a background absorption in the spectral range of 0.1 to 3.0 eV due to defects with increasing O/Ni ratio. Ono *et al* [33] studied the relationship of the transmittance and electrical properties of NiO and suggested that $V_{Ni}$-induced *p-d* charge transfer transitions may have led to the darker color of the sputter-deposited thin films. However, a critical study on the optical properties, including the complex dielectric functions and optical constants have not been explored for these NiO systems with Ni vacancies showing *p*-type character. Hence, detailed knowledge of the effects of native defects such as $V_{Ni}$ and $V_O$ on the optical properties of NiO are still lacking.

In this study, a comprehensive investigation of the correlation between vacancy defects and the optical behavior and electronic properties of NiO was performed using Spectroscopic Ellipsometry (SE) and X-ray Photoemission Spectroscopy (XPS) on NiO thin films sputter deposited in stoichiometric, Ni-rich and O-rich environments. We note that when NiO is grown in an O-rich or Ni-rich environment, aside from the favorable formation of Ni vacancies or O vacancies respectively, it is also possible that other native defects such as interstitials and complexes would be formed. However, in NiO grown in an O-rich environment, it has been established that the formation energy of Ni vacancies is much lower than that of other defects[29]. Hence it is believed that they are responsible for its observed *p*-type behavior. Thus, following previous reports that vacancy type defects have lower formation energies[20,29,34], interstitial type defects are not considered in this work. In the following, Ni-rich and O-rich NiO thin films are referred to as NiO:$V_O$ (NiO with O vacancies) and NiO:$V_{Ni}$ (NiO with Ni vacancies), respectively. The term "stoichiometric" here refers to films sputtered from a NiO ceramic target using Ar gas only. Standard Spectroscopic ellipsometry (SE) was used to obtain the complex dielectric functions, optical constants and absorption coefficient of NiO samples with different vacancy defects [35]. SE is a well-known optical technique for determining the dielectric function and hence, the optical constants of thin films with high precision. It has been used to study the optical properties of several transparent metal oxides[36,37]. In addition, first-principles DFT + *U* calculations of the optical properties and electronic structures of NiO with Ni and O vacancies are also carried out and these computed results are directly compared to the experimental data.

# EXPERIMENTAL AND COMPUTATION PROCEDURE

## Materials Synthesis and characterization

Stoichiometric NiO, Ni-rich NiO (NiO:$V_O$), and O-rich NiO (NiO:$V_{Ni}$) thin-film samples were synthesized using a radio-frequency magnetron sputtering system at a substrate temperature of 280°C. Ni metal target and NiO ceramic targets were used during the sputtering process to deposit samples on soda-lime glass. The oxygen flow ratio, $r(O_2) = f(O_2)/[f(Ar)+f(O_2)]$ in the sputtering gas was controlled by separately regulating the flow rate of pure Ar and $O_2$ during deposition. Stoichiometric NiO samples were deposited in a pure Ar environment ($r(O_2) = 0$) from a stoichiometric NiO (99.99%) ceramic target while O-rich NiO:$V_{Ni}$ samples were deposited with $r(O_2) = 4\%$. Ni-rich NiO:$V_O$ samples were deposited by co-sputtering a Ni metallic target and the NiO target. Ni content is controlled by the DC sputtering power for the metal Ni target.

Electrical properties of deposited films were investigated by Hall-effect measurements in the *van der Pauw* geometry with a magnetic field of 0.6T using a commercial (Ecopia HMS-5500) system. NiO:$V_{Ni}$ samples showed *p*-type conductivity (with resistivity, $\rho \sim 3.6$ Ω-cm) while the $\rho$ of NiO:$V_O$ and stoichiometric NiO were too high to measure with our Hall system ($\rho > 10^4$ Ω-cm). Because of the low carrier mobility in NiO:$V_{Ni}$ (<1 cm$^2$/Vs), Hall effect measurement cannot reliably determine the conductivity type of the sample at room temperature. The conductivity type was verified by thermopower measurements using a commercial MMR SB1000 Seebeck system. We confirmed that NiO:$V_{Ni}$ films are *p*-type with a positive Seebeck coefficient of +84 $\mu$V/K.

Film stoichiometry and thickness were characterized by Rutherford backscattering (RBS) using a 3.04 MeV He$^{++}$ beam, and the spectra were analyzed using the SIMNRA software[38]. We note that while RBS can be sensitive to high Z impurities to the part per million levels, the determination of the O stoichiometry in NiO to within 1% precision is challenging. Hence only a qualitative measure of the O stoichiometry in NiO samples is given here by RBS. Film thicknesses in the range of 90-150 nm were obtained by RBS and were confirmed by spectroscopic ellipsometry thickness fitting. Grazing incidence x-ray diffraction (GIXRD) with an incidence angle of 1° was used to characterize the film structure. The surface morphology of all films was studied by Atomic Force Microscopy (AFM).

The optical properties of the films were studied by standard (isotropic) Spectroscopic Ellipsometry (SE) in the spectral range of 0.73eV to 6.5eV using a rotating-compensator instrument (J. A. Woollam, M-2000). The angle of incidence was varied from 55° to 75° with an increment of 10°. To eliminate back-side reflection, the glass substrate backside was taped with a translucent plastic tape[39]. The electronic structure of deposited films were obtained by high-resolution x-ray photoelectron spectroscopy (XPS) using a monochromatic Al K$\alpha$ X-ray source (1.487keV) and emitted photoelectrons were collected and analyzed using a concentric hemispheric analyzer system. In XPS measurements, the core level electron binding energies were referenced to the adventitious C 1$s$ at 284.8 eV to correct for electrostatic charging effects for insulating films.

**First Principle Calculation**

First-principles calculations were performed using density functional theory (DFT)[40,41] as implemented in *Vienna Ab Initio Simulation Package (VASP)*[42]. We used the Perdew-Burke-Ernzerhoff (PBE) functional[43] for the exchange and correlation, an effective Coulomb interaction $U$ (DFT + $U$) to account for electron correlations and exchanges of the valence $d$ shell electrons. We select for the Ni $d$ shell the Coulomb $U$ and exchange energy parameter $J$ of 5.10 and 0.95 eV, respectively, using the rotationally invariant method of Liechtenstein *et al.*[44]. The choice of the effective interaction value is motivated by the experiment as discussed below. The unit cell of NiO as shown in Figure 1(a) is rock-salt with a rhombohedral symmetry due to the type-II antiferromagnetic order along the [111] direction. We used a 108-atom rhombohedral supercell to model the defects by randomly removing a pair of Ni or O atom, i.e., one vacant site in each of the magnetic sublattices using the special quasirandom structures[45] with the mcsqs utility in the Alloy Theoretic Automated Toolkit[46]. The choice of a vacant site on each of the magnetic sublattice at a time is to maintain the antiferromagnetic symmetry of the parent material[47]. The resulting crystal structures for both Ni and O vacancy contain isolated defect sites that are more than 15 Å apart. Herein, our simulation is for the lowest possible defect concentration for the 108 atoms supercell, which corresponds to 3.70% of the total Ni or O sublattice sites. To optimize the crystal structures, we employed a $\Gamma$-point sampling of the reciprocal space to relax the structures until the energy (charge) was converged to within ~$10^{-3}$ ($10^{-6}$) eV and the forces dropped to $10^{-3}$ eV/Å. The relaxed structure shows that nearest-neighbor oxygen atoms around the Ni defect sites moved outward by ~0.19 Å. On the other hand, the Ni atoms in the immediate surroundings of the O atom defect sites relaxed inward by ~0.11 Å. These two opposing relaxation patterns are due to the electrostatic interactions between the defect sites and proximity atoms, which induced spatial distribution of the defect levels. For example, the effective charge of $V_O$ is positive, the same sign as the neighboring Ni atoms and will repel each other leading to the relaxation pattern described above. To obtain the electronic structures, we employed a $3 \times 3 \times 3$ k-point grid to represent the reciprocal space and were performed in the antiferromagnetic phase using a collinear spin-polarized approach with a kinetic energy cutoff of 550 eV for the planewave basis set. The optical spectra were obtained using independent particle approximation as implemented in the VASP code.

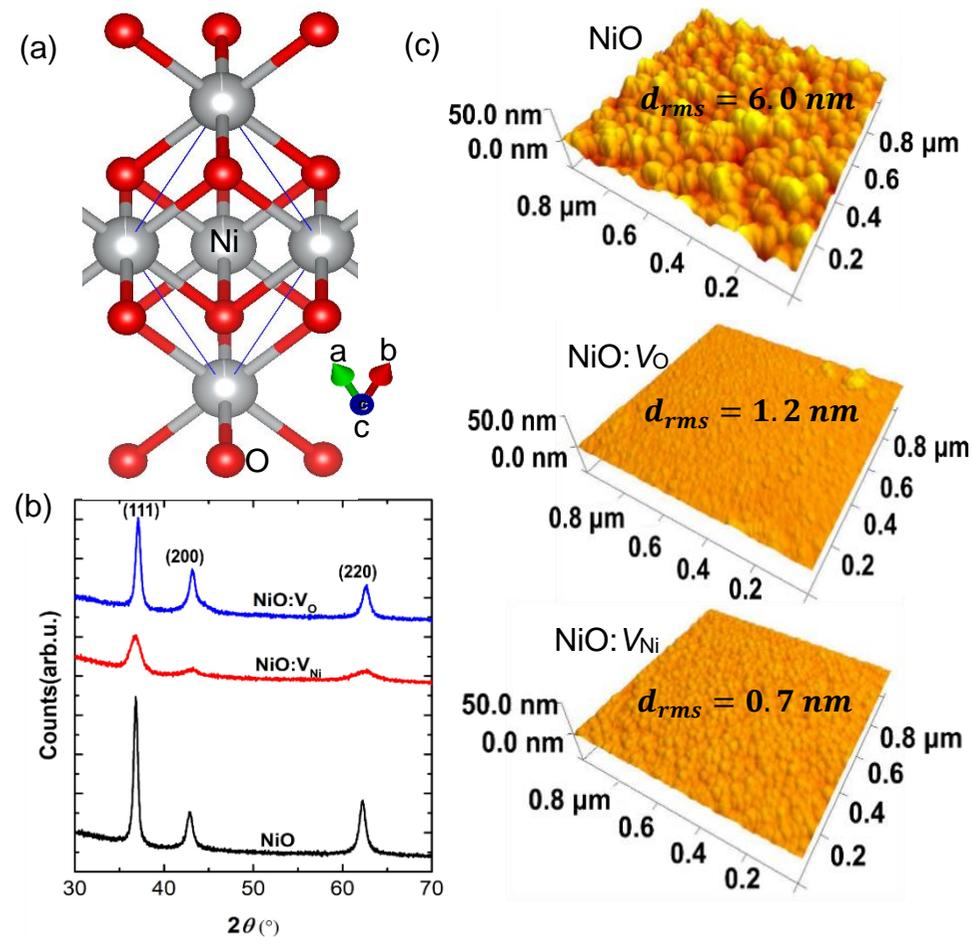

**FIG 1**. (a) The unit cell (denoted with a solid blue line) of the antiferromagnetic NiO crystal structure. (b) X-ray diffraction patterns from NiO, NiO:$V_{Ni}$ and NiO:$V_O$ thin films (c) Atomic force microscopy (AFM) images of the samples, the root-mean-square roughness ($d_{rms}$) obtained from each AFM image is specified.

## Spectroscopic Ellipsometry (SE) Analysis

The dielectric functions ($\varepsilon = \varepsilon_1 + i\varepsilon_2$) of NiO, NiO:$V_{Ni}$ and NiO:$V_O$ films were evaluated by spectroscopic ellipsometry (SE). In the SE analysis, we assumed an optical model comprising of a surface roughness/thin films/glass substrate structure. The refractive index ($n$) and extinction coefficient ($k$) of the glass substrate were obtained as described previously[36]. The Bruggeman effective medium approximation (EMA) with a 50:50 vol% mixture of the bulk layer and void was used to model the optical properties of the surface roughness layer[35]. The Tauc-Lorentz oscillator model was used to fit the amplitude ratio, $\Psi$ and the phase difference $\Delta$ spectra for stoichiometric NiO and NiO:$V_O$ films, which have no significant absorption for $E \leq E_g$. In this model, the imaginary part of the dielectric function, $\varepsilon_2(E)$ is expressed as a product of the Tauc optical gap and the Lorentz model.

$$\varepsilon_2(E) = \begin{cases} \dfrac{AE_0\Gamma(E-E_g)^2}{[(E^2-E_0^2)^2+\Gamma^2 E^2]}\dfrac{1}{E} & (E > E_g) \\ 0 & (E \leq E_g) \end{cases} \quad (1)$$

where $A$, $E_0$, $\Gamma$ and $E_g$, represent the Lorentz amplitude parameter, center transition energy, broadening parameter and Tauc optical gap, respectively[48]. The Tauc Lorentz $\varepsilon_1(E)$ spectra is then obtained using the Kramers-Kronig relation. At high energies, $\varepsilon_\infty$ describes a constant, energy independent contribution to $\varepsilon_1(E)$. Hence, in the Tauc-Lorentz model, the dielectric function is described by five parameters ($A$, $E_0$, $\Gamma$, $E_g$ and $\varepsilon_\infty$). Two Tauc-Lorentz dielectric functions (Table 1) were used to accurately fit $\Psi$ and $\Delta$ of NiO and NiO:$V_O$ samples. The fitted and experimental $\Psi$ and $\Delta$ spectra of NiO are shown in Figure 2. The surface roughness, $d_s$ of 4.2 $\pm$ 0.1 nm obtained from the SE analysis of the NiO film is in close agreement with the value of the roughness $d_{rms}$ = 6.0 nm observed in AFM (Figure 1(c)), confirming the validity of the SE analysis. Also, the mean-square-error (MSE) which is a measure of goodness of fit has a low value of 2.80. The MSE sums the differences between the measured data and the model generated data over all the measurement wavelengths of the ellipsometric $\Psi$ and $\Delta$ spectra; the MSE is minimized by adjusting the the fit parameters using the Levenberg-Marquardt non-linear regression algorithm. The Tauc-Lorentz fitting parameters for NiO and NiO:$V_O$ are summarized in Table 1.

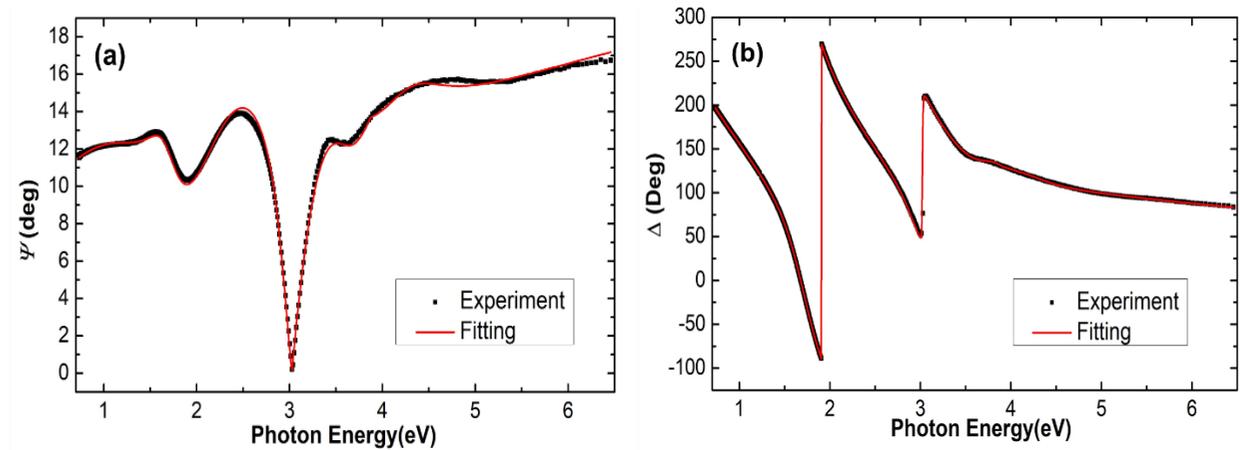

**FIG 2**: The measured (black square) and fitted (red solid line) SE spectra of (a) amplitude ratio ($\Psi$) and (b) phase difference ($\Delta$) for NiO thin film measured at an incident angle of 65°.

For O-rich NiO samples (NiO:$V_{Ni}$), the dielectric function was determined from fittings of the SE data using the Cody-Lorentz(C-L) model and results are shown in Figure S1(Supplementary Material). Gaussian and Lorentz oscillators were also applied in modeling the system to improve the fitting of the dielectric function, especially at higher energies above the bandgap. Ferlauto *et. al.*,[49] developed the Cody-Lorentz model to describe amorphous materials, showing a significant sub-gap absorption and absorption on the band tail regions. Though the Tauc-Lorentz (T-L) model has been used widely to model dielectric functions of transparent metal oxide materials, it assumes $\varepsilon_2 = 0$ for energies $E \leq E_g$ as shown in Eq. 1, and hence may not fully account for the absorption behavior below the bandgap. On the other hand, the Cody Lorentz dielectric function model can provide an improved description of the absorption coefficient in the region below the bandgap by including parameters to model absorptions at energies below $E_g$. In contrast to the other samples, NiO:$V_{Ni}$ showed higher sub-gap absorptions, which necessitated the use of the Cody-Lorentz model to enable a more accurate fitting of absorptions below the bandgap. Like the Tauc-Lorentz, the Cody-Lorentz oscillator defines an energy bandgap and the Lorentzian absorption peak parameters. However, for the region just above $E_g$, the Cody-Lorentz absorption formula is given by $\varepsilon_2(E) \propto (E-E_g)^2$ while the Tauc-Lorentz absorption uses the dependence $\varepsilon_2(E) \propto [(E - E_g)^2/E^2]$. The details of the Cody-Lorentz model fitting parameters are described in the Supplementary Material. The extracted fitting parameters for the Cody Lorentz models are also shown in Table 1. In Table 1, for the NiO and NiO:$V_O$ fitting using the Tauc-Lorentz oscillator, the first Tauc-Lorent oscillator, T-L1, is used to model interband absorption close to the absorption edge of the material. The second Tauc-Lorentz oscillator T-L2 models other absorption at higher energies above the absorption edge associated with the electronic interband transitions.

**Table 1**: The peak fitting parameters extracted from the dielectric function of thin films

| Material | Peak | $A$ | $E_0$ (eV) | $\Gamma$ (eV) | $E_g$ (eV) | $\varepsilon_1(\infty)$ | $E_u$ (eV) | $E_t$ (eV) | $E_p$ (eV) |
|---|---|---|---|---|---|---|---|---|---|
| **NiO** | T-L1 | 39.778 | 4.048 | 1.354 | 2.679 | 2.251 | - | - | - |
|  | T-L2 | 22.940 | 6.530 | 5.164 | 1.316 | 0 | - | - | - |
| **NiO:$V_O$** | T-L1 | 48.606 | 4.138 | 1.480 | 2.803 | 2.283 | - | - | - |
|  | T-L2 | 18.970 | 7.935 | 5.426 | 0.675 | 0 | - | - | - |
| **NiO:$V_{Ni}$** | C-L1 | 36.623 | 3.074 | 7.000 | 3.074 | 2.122 | 0.549 | 0.840 | 0.835 |
|  | L1 | 1.504 | 2.386 | 4.987 | - | - | - | - | - |
|  | G1 | 2.186 | 8.399 | 4.120 | - | - | - | - | - |

T-L: Tauc-Lorentz, C-L: Cody Lorentz, L: Lorentz, G: Gaussian

## RESULTS AND DISCUSSION

### Structural Characterization

XRD patterns of NiO, NiO:$V_O$ and NiO:$V_{Ni}$ thin films shown in Figure 1(b) reveal that all films are polycrystalline, showing diffraction peaks from the rock-salt (111), (200), and (220) planes. A decrease in the crystallinity is evident in films with defects. The significant decrease in the XRD peak intensity per unit film thickness and increase in the FWHM for the NiO:$V_{Ni}$ and NiO:$V_O$ thin films suggest smaller grain size in the films and are likely due to structural disorders. In particular, the NiO:$V_{Ni}$ film shows much broader diffraction peaks, indicating a much smaller grain size of

~9 nm compared to the ~18 nm gain size for the "stoichiometric" NiO film. The atomic force microscopy (AFM) results presented in Figure 1(c) show that the "stoichiometric" film has a root-mean-square (rms) roughness of ~ 6.0 nm compared to the rms roughness of 1.2 and 0.7 nm for NiO:$V_O$ and NiO:$V_{Ni}$, respectively. The smoother film surface of the films with vacancy defects is consistent with their smaller grain size revealed by XRD.

**Optical properties**

The dielectric functions of the deposited thin films obtained by model fittings of the SE results are compared with results computed using the calculated electronic wave functions and energies. The measured ((a) and (b)) and calculated ((c) and (d)) real $\varepsilon_1$ and imaginary $\varepsilon_2$ parts of the complex dielectric functions as a function of the photon energy for the three different NiO samples are shown in Figure 3. We note that the computed spectra for NiO show similar features as the experimental spectra up to a photon energy of ~4.30 eV, but they show a significant deviation especially in $\varepsilon_1$ for higher photon energies. This deviation could be due to several sources in both the computation and experimental techniques. For example, the computed spectra in NiO show more structures due to the small broadening parameter ~$10^{-3}$ used in our calculations. The small grain size of the films (~ < 20 nm) may play a role in broadening the experimental spectra, rendering these sharp features not observable. The deviation could also be attributed to the inability of DFT-based optical calculations to properly describe high photon energy states[50]. The corresponding optical constants, *n* and *k* obtained from ellipsometry fitting are shown in Figure S2(Supplementary Material). Table 2 compares the refractive index, *n* at 2 eV for NiO obtained in this work and values reported previously in the literature.

**Table 2**: Comparison of reported refractive index of NiO at 2eV

| Refractive index at 2eV | Method/Model | Ref |
| --- | --- | --- |
| 2.37 | SE (Tauc-Lorentz Model) | 51 |
| 2.40 | SE (Tauc-Lorentz Model) | 52 |
| 2.33 | Reflectance | 53 |
| 2.38 | SE | 54 |
| 2.40 | SE | 55 |
| 2.40 | SE (Tauc-Lorentz Model) | This work |

Focusing first on the stoichiometric NiO sample, the dispersive part of the experimental dielectric function is presented in Figure 3(a). The photon-energy dependence of $\varepsilon_1$ is typical of most transparent metal oxides with a peak structure at ~3.6 eV. In the computed spectrum (Figure 3(c)), this peak appears at ~3.2 eV. This is in agreement with previously reported calculations for NiO[56]. The experimental $\varepsilon_2$ spectrum for stoichiometric NiO show negligible features between 1 eV to 3 eV with a characteristic peak at ~ 4.0 eV, which occurs at around 3.90 eV in the calculated spectrum. This peak is related to the absorption edge in NiO. The reported experimental NiO $\varepsilon_2$ spectrum is in good agreement with previous reports[42-44].

To explore the impact of defects on the optical properties, we compare in Figure 3 the experimental and computed NiO $\varepsilon_1$ and $\varepsilon_2$ spectra with those from NiO:$V_{Ni}$ and NiO:$V_O$. In Figure 3(a), the experimental $\varepsilon_1$ spectra of the NiO:$V_O$ sample closely follows that of the stoichiometric NiO sample. However, a slight increase in the NiO:$V_O$ spectrum at lower photon energy is observed compared to NiO. On the other hand, the $\varepsilon_1$ spectrum of NiO:$V_{Ni}$ shows a stronger deviation from that of NiO. A significant increase in $\varepsilon_1$ is observed for the NiO:$V_{Ni}$ at photon energies below 2.3 eV. The computed results in Figure 3(c) also show similar $\varepsilon_1$ features for the NiO:$V_O$ and stoichiometric NiO sample which are very different from that for the NiO:$V_{Ni}$ sample with the same defect concentration. Due to the presence of O-vacancies in the NiO:$V_O$ slight differences in the computed $\varepsilon_1$ spectrum for NiO:$V_O$ are observed compared to NiO spectrum. A comparison of the computed $\varepsilon_1$ spectrum for NiO:$V_O$ with previously reported calculation shows that the peak position of our calculated data shifts slightly to ~3.5 eV compared to ~ 4 eV shown in the previous calculation. This slight discrepancy may be due to the different *U*-value employed in the previous calculation [57]. A similar feature at ~ 1.9 eV observed in the calculation by Peterson *et.al.,* can be observed in our data, though, with smaller intensity likely due to differences in defect concentration in the calculations [57].

While the computed and experimental $\varepsilon_1$ for NiO and NiO:$V_O$ samples show similar features, we observe significant deviations between the computed and experimental $\varepsilon_1$ in NiO:$V_{Ni}$. In particular, we note that the peak around ~ 3.0 eV is not observable in the calculated data. A possible reason may be that the vacancy concentration used in the calculation is larger than that in the grown samples. It is also possible that the independent particle approximation used in the calculation did not properly capture the low-photon spectra for NiO:$V_{Ni}$.

The $\varepsilon_2$ spectra also reveal the different behavior of the studied native defects. The experimental and calculated $\varepsilon_2$ spectra are shown in Figure 3(b) and (d), respectively. A slight blue-shift of the fundamental absorption peak at ~4.0 eV and noticeable non-zero absorptions below ~3.0 eV are observed in the experimental NiO:$V_O$ compared to stoichiometric NiO. The sub bandgap absorption can be attributed to $V_O$ defect states in the bandgap. In contrast, the experimental $\varepsilon_2$ spectrum for NiO:$V_{Ni}$ exhibits a significant red-shift of the band-edge, a larger non-zero absorption below ~3.0 eV, and a reduced amplitude of the characteristic absorption peak at ~4.0 eV.

In the computed data in Figure 3(d) the $\varepsilon_2$ spectrum for NiO:$V_O$ closely follows that of the stoichiometric NiO but deviates from that for NiO:$V_{Ni}$. We notice the emergence of a small local maximum around ~2 eV (1 eV) in the computed $\varepsilon_2$ spectrum for NiO:$V_O$ (NiO:$V_{Ni}$) that is not resolved in our experimental data. Petersen *et al.* reported a similar feature at ~ 2 eV albeit with higher intensity in their NiO with O vacancies calculation[57]. We believe that because of the small amplitude of this structure, it might have been masked by the broadening parameter used in obtaining experimental data, which instead gives rise to a broad shoulder around 1.3 eV in the experimental $\varepsilon_2$ spectra of NiO:$V_{Ni}$ (Figure 3(b)).

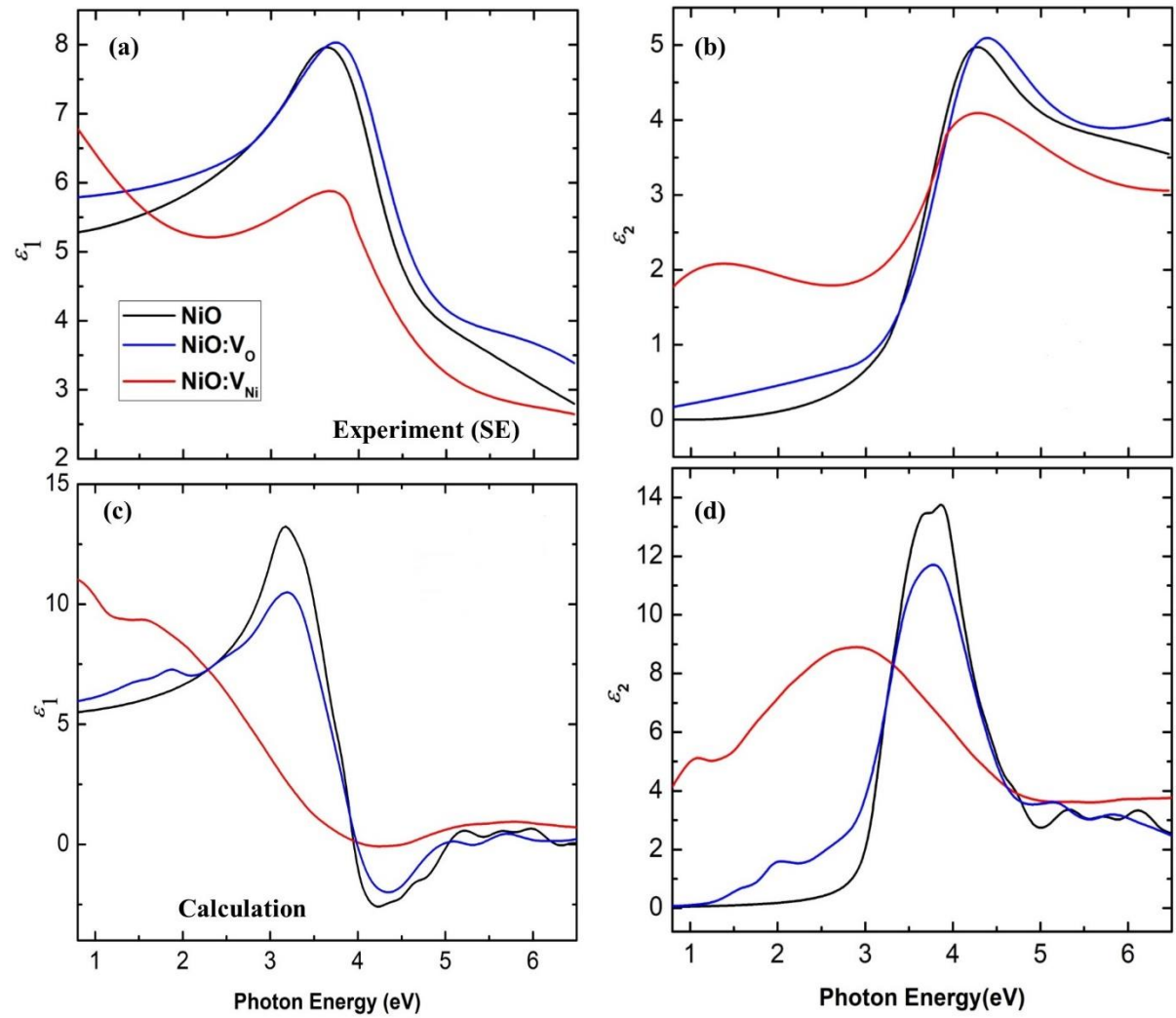

**FIG 3**: Photon energy dependence of the real $\varepsilon_1$ (a) and imaginary $\varepsilon_2$ (b) parts of the complex dielectric function for NiO, NiO:$V_O$ and NiO:$V_{Ni}$ extracted from the SE analysis. The corresponding calculated spectra are presented in (c) and (d), respectively.

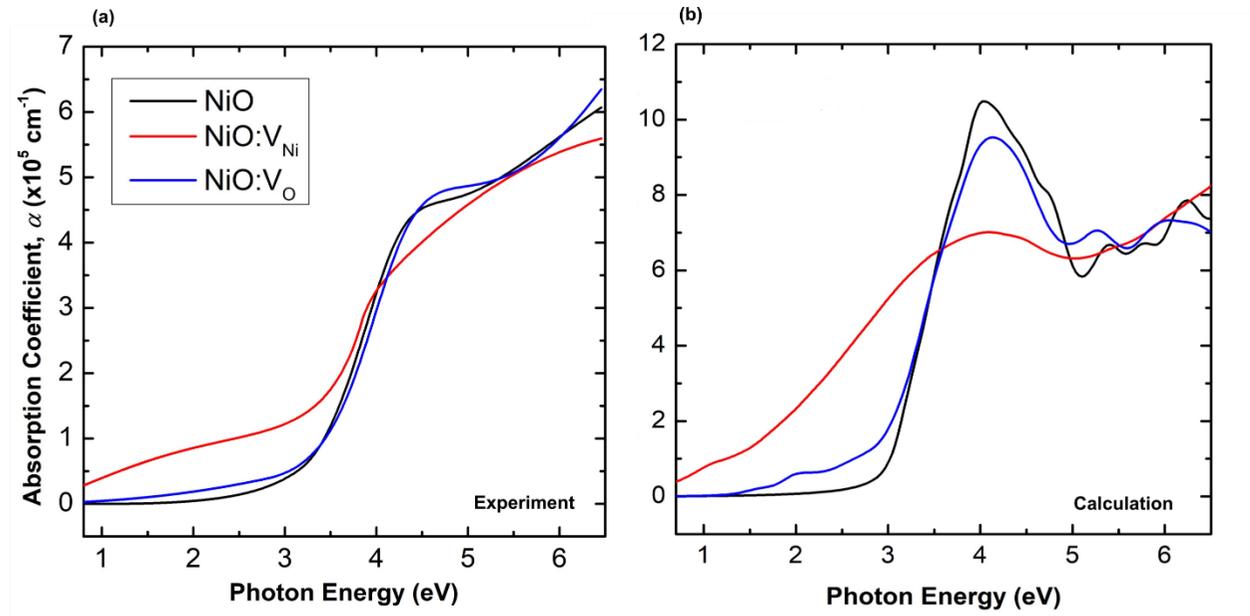

**FIG 4**: Absorption coefficient spectra of NiO, NiO:$V_O$ and NiO:$V_{Ni}$ samples obtained from (a) spectroscopic ellipsometry and (b) *ab initio* calculations.

Figures 4(a) and (b) show absorption coefficient $\alpha$ obtained from the experimental and calculated dielectric functions respectively for the various samples. The experimental α presented in Figure 4(a) show an onset of absorption around ~3.5 eV for NiO and a slightly lower value of ~3.4eV for NiO:$V_{Ni}$. Below the absorption edge in the experimental absorption spectra, stoichiometric NiO and NiO with native vacancies show very different behaviors. Sub bandgap absorption due to defect states are observed in NiO with vacancy defects. Figure 4(a) shows that NiO:$V_{Ni}$ has a larger subgap absorption compared to the NiO:$V_O$ sample. For example, at a photon energy of 2.5 eV, α is ~$1\times10^5$ and $3\times10^4$ cm$^{-1}$ for NiO:$V_{Ni}$ and NiO:$V_O$, respectively.

The absorption edge of NiO obtained from the calculated absorption coefficient in Figure 4(b) is slightly lower at ~ 3.1 eV compared to the experimental value due to the choice of our *U*-value in the calculation. Similar changes in the experimental absorption spectra for the samples with vacancy defects are seen in the calculated data. While stoichiometric NiO has no significant absorption below the onset for both experimental and calculated spectra, NiO:$V_O$ shows some absorption features between 2-3 eV before the onset of absorption at ~3.5 eV for the experimental and 3.1 eV for the calculated spectra. NiO:$V_{Ni}$ on the other hand shows a large sub-gap absorption below the onset at 3.4 eV in the experimental data while in the calculated data a large sub-gap absorption is also seen which increases with photon energy. Subgap absorption similar to our experimental data has been reported previously for p-type NiO grown in O-rich environment[31,58]. These changes in the absorption spectra for the NiO:$V_{Ni}$ is attributed to the action of acceptor-type $V_{Ni}$ producing Ni$^{3+}$ and a free hole by trapping an electron from Ni$^{2+}$ leading to an increase in charge transfer transitions in the visible range[33].

The high sub-bandgap absorption coefficient observed in the NiO:$V_{Ni}$ thin film leads to a large reduction in visible light transparency of the NiO:$V_{Ni}$ thin films. Figure S3(Supplementary Material) shows the corresponding measured transmittance, $T$ spectra for the NiO, NiO:$V_O$ and NiO:$V_{Ni}$ films. The transmittance of the stoichiometric NiO is ~75% in the visible range. For the samples with vacancy defects, $T$ decreases to ~60% for NiO:$V_O$ and ~40% for NiO:$V_{Ni}$. This decrease in transmission intensity is less profound compared to those presented by Ono *et. al*[33]. We note that in Ono's work, NiO films were sputtered in pure O$_2$ plasma while the O-rich NiO films we synthesized were sputtered in a Ar+O$_2$ mixed gas with only 4% O$_2$. This suggests that NiO films in Ono's work were much more O-rich with a much high $V_{Ni}$ concentration. Similar reduced transmittance in the visible regime for *p*-type O-rich NiO and NiO based alloys were also reported in previous studies[30,59,60]. To further investigate the nature of the high sub-gap absorption of the films with vacancy defects, we characterize the electronic structures with first-principles calculations and X-ray photoelectron spectroscopy measurements.

### Electronic structure: DOS and XPS

To investigate the electronic properties, we obtain the valence band spectra from XPS measurements while the electronic structures were obtained from first-principles calculations. Figure 5 shows the spin-polarized density of states (*DOS*) for stoichiometric and disordered NiO calculated using the GGA+U functional. The value of the bandgap for stoichiometric NiO with our calculations using $U$ of 5.1 eV is ~2.80 eV. The calculated NiO electronic properties and the predicted $E_g$ are in good agreement with previous calculations with similar $U$-value[61,62]. As the $U$-value is increased, the calculated $E_g$ is increased, e.g., $E_g$~3.0 $\pm$ 0.2 eV for $U$~6.0-6.5 eV has previously been reported[63–65]. A larger value of $U$~8 eV in our calculation can lead to a larger bandgap, $E_g$~4.2 eV, which is closer to the generally accepted experimental value in the range of 3.6 – 4.3 eV[53,66,67]. However, the agreement of the absorption spectra obtained with such a large $U$ value with experimental data is rather poor. Overall, we do not expect the application of the GGA + $U$ to the Ni $d$ shell to obtain the accurate band gap value for NiO; as it is well known that band gaps are significantly underestimated in semi-local functionals even for materials without transition metal $d$ states such as Si and Ge. Nevertheless, our calculated total density of states for NiO shown in Figure 5(a) agree with the results reported in the literature[29,61,68].

Modeling defects in an antiferromagnetic system requires some care, especially at the dilute limit. Antiferromagnetic systems like NiO have symmetric spin-up and spin-down sublattices as such, there is no sublattice preference for defect levels, i.e., the majority spin on a given sublattice is the minority spin on the other sublattice, and vice versa. Hence, a single Ni or O vacancy, especially at the dilute limit will break the symmetry between the two magnetic sublattices in NiO. Our calculations show that single-atom vacancy leads to half-metallicity, while current experiments show robust antiferromagnetic insulating state. The observation of half-metallicity by a one-atom vacancy in computations has been shown in previous studies[28,47]. We can ascertain the stability of the half-metallicity through random defect modeling and spin-polarized photoemission measurement, which will be the focus of a future study. At the dilute limit, one would expect Ni/O vacancy levels to be randomly distributed without sublattice preference; this could preserve the antiferromagnetic configuration of pristine NiO. In our defect modeling, we have adopted one vacant site in each of the magnetic sublattices to preserve the antiferromagnetic symmetry of NiO.

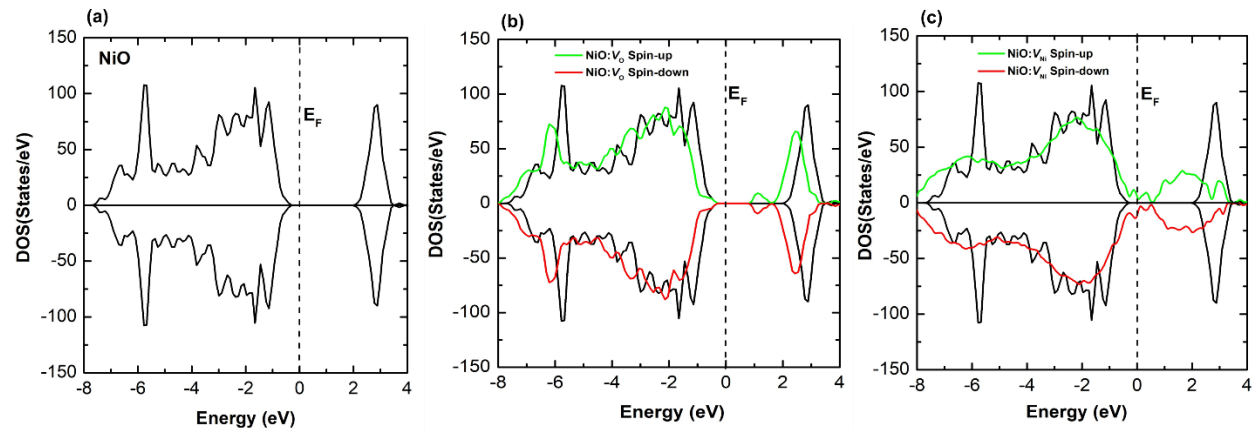

**FIG 5:** Calculated electronic structures of (a) stoichiometric NiO and NiO:x with 3.7% vacancy defects for (b) x = $V_O$, (c) x = $V_{Ni}$. The vertical dashed line is the Fermi level, which is set at zero energy.

We present in Figure 5(b) and (c) the calculated electronic structures of NiO with native O and Ni vacancy defects, respectively. One can see that the *DOS* for both spin channels are identical, confirming the stability of antiferromagnetic states. Distinct features, especially in-gap levels appear in both NiO:$V_O$ and NiO:$V_{Ni}$. Overall, the defect levels have more impact on the conduction band states. From the projected *DOS* [Figure S4(Supplementary Material)], the defect-induced in-gap levels above the Fermi level in the NiO:$V_O$ is dominated by the Ni-3$d$ states. In contrast, the defect-induced in-gap states in NiO:$V_{Ni}$ are dominated by a strong hybridization between Ni-3$d$ and O-2$p$ states. Interestingly, more pronounced in-gap states are visible in the NiO:$V_{Ni}$ compared to the NiO:$V_O$ with the same defect concentration (3.7%). Specifically, the NiO:$V_{Ni}$ system leads to a half-metallic solution, in basic agreement with the works of Ködderitzsch *et al*,[47] and Park *et al*,[28] which is due to spin polarization of the atoms in the proximity of the vacant Ni sites.

Figure 6 shows the XPS spectra close to the Fermi level for the NiO, NiO:$V_O$ and NiO:$V_{Ni}$ thin films. The calculated *DOS* of NiO convoluted with a Gaussian broadening, $\gamma$= 0.8 using the Galore code[69] is also shown for comparison. Two peak energy regions can be distinguished in the calculated *DOS*, between -1.0 eV to -3.0 eV and -5.0 eV to -8.0 eV. From the PDOS [in Figure S4(Supplementay Material)] pronounced Ni-3$d$ orbital states are concentrated in the energy -1.0 eV to -3.0 eV range with a peak at ∼ -2.2 eV, while in the energy range between -5.0 eV to -8.0 eV, O 2$p$ states also have significant contribution to the spectrum. The experimentally obtained XPS data agree closely with the broadened calculated NiO DOS. All three valence band spectra of NiO, NiO:$V_{Ni}$ and NiO:$V_O$ show a peak at ∼ -2 eV, corresponding to predominant contributions from the Ni 3$d$ states, and consistent with the peak position of the calculated DOS. However, features due to O 2$p$ states around -5.0 eV to -8.0 eV are less pronounced. This may be attributed to the Al K$\alpha$ (1486.7eV) source used in the measurement which has a more favorable photoionization cross-section for the $d$ electrons. A similar observation has been made for Cu based oxide systems[70].

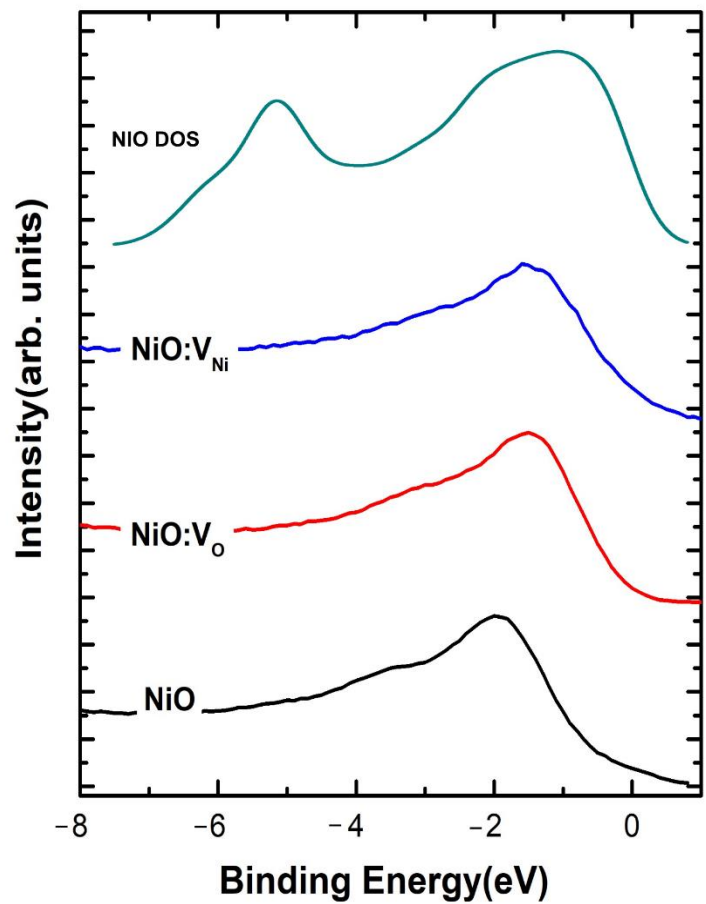

**FIG 6**: Experimental XPS valence-band spectra of NiO, NiO:$V_O$ and NiO:$V_{Ni}$ thin films near the Fermi level. The calculated NiO total density of state convoluted with a Gaussian broadening of 0.8 using the Galore code is also shown for comparison (top spectrum, labelled NiO DOS)

The observed in-gap states in both NiO:$V_{Ni}$ and NiO:$V_O$ systems confirm the sub-gap absorption in our optical spectra. Also, the half-metallic solution for NiO:$V_{Ni}$ supports the rather high *p*-type conduction observed in NiO:$V_{Ni}$ thin films as compared to stoichiometric NiO and NiO:$V_O$ thin films. Hence, achieving *p*-type conductivity in NiO by increasing $V_{Ni}$ will necessarily degrade its transparency in the visible range due to this formation of half metallicity in the electronic structure. Therefore, optoelectronic applications requiring high transparency and good *p*-type conduction in NiO would require a systematic combination of small amounts of $V_{Ni}$ native defects with extrinsic doping to reduce the gap states introduced in the material while maximizing conductivity. This alternative route to simultaneously optimize *p*-type conductivity and transparency of NiO would be a subject of further study. Elements such as Li, Cu, or Ag have been studied as acceptor dopants

Journal of Applied Physics — ACCEPTED MANUSCRIPT
This is the author's peer reviewed, accepted manuscript. However, the online version of record will be different from this version once it has been copyedited and typeset.
PLEASE CITE THIS ARTICLE AS DOI: 10.1063/5.0021650
AIP Publishingin NiO [31,71,72]. A systematic investigation of the optical properties of NiO combining $V_{Ni}$ and these acceptor elements may pave the way to realizing transparent and conducting *p*-type NiO for transparent electronics applications.

## CONCLUSION

We have experimentally investigated changes in the complex dielectric function, absorption coefficient, and electronic structure of NiO due to vacancy defects, and results are supported by computational studies. The complex dielectric functions and optical constants of stoichiometric NiO and NiO with vacancy defects were obtained by fitting the amplitude ratio and phase difference of spectroscopic ellipsometry measurements using oscillator models. We show that the optical constants of *p*-type O-rich NiO (NiO:$V_{Ni}$) deviate substantially from the stoichiometric NiO. By comparing the effects of native vacancies in NiO, we establish that $V_{Ni}$ present in NiO grown under an O-rich environment leads to a significant sub-bandgap absorption as compared to $V_O$ in NiO grown under Ni-rich condition. The calculated density of states confirms a high concentration of gap states in NiO:$V_{Ni}$, resembling a half-metallic system as compared to a smaller subgap state in NiO:$V_O$. Hence, this significant density of gap states in NiO:$V_{Ni}$ arising from the introduction of $V_{Ni}$ acceptors in NiO can explain the high absorption in the visible range for this material. Interestingly, while Ni 3*d* states are responsible for the gap states in NiO:$V_O$, both Ni 3*d* and O 2*p* contribute to those in NiO:$V_{Ni}$. Our results establish the relation between vacancy native defects and the optical and electronic properties of NiO. Though native $V_{Ni}$ is an effective acceptor leading to *p*-type conductivity in NiO, they also give rise to a high density of gap states in the electronic structure and hence degrade the visible transparency of the material. Therefore, for applications as a *p*-type transparent conductor in solar cells and other optoelectronic devices, it is essential to also explore the combination of extrinsic doping with native vacancy defect control in NiO.

## SUPPLEMENTARY MATERIAL

See Supplementary material for details of Spectroscopic ellipsometry data analysis using Cody-Lorentz model for the NiO:$V_{Ni}$ thin film. Plots of the optical constants of the thin films, transmission intensity and the calculated partial density of states of the the systems are also shown in the Supplementary material.

## DATA AVAILABILITY

The data that supports the findings of this study are available within the article and its supplementary material.

## ACKNOWLEDGEMENTS

This work was supported by the General Research Fund of the Research Grants Council of Hong Kong SAR, China, under Project No. CityU 11267516 and CityU-SRG 7005106. C.P.L acknowledges the support of the start-up fund from Shantou University under Project No. NTF18027, Guangdong Basic and Applied Basic Research Foundation (Project No. 2020A 1515010180), the Major Research Plan of the National Natural Science Foundation of China (Project No. 91950101), and the Optics and Photoelectronics Project (No. 2018KCXTDO11). Supercomputer and computational resources were provided by the Lehigh University (LU) High-Performance Computing Center. C.E.E acknowledges LU start-up and summer research fellowship. K.O.E was supported by the Hong Kong Ph.D. Fellowship (PF16-02083) Research Grants Council, University Grants Committee, Hong Kong.

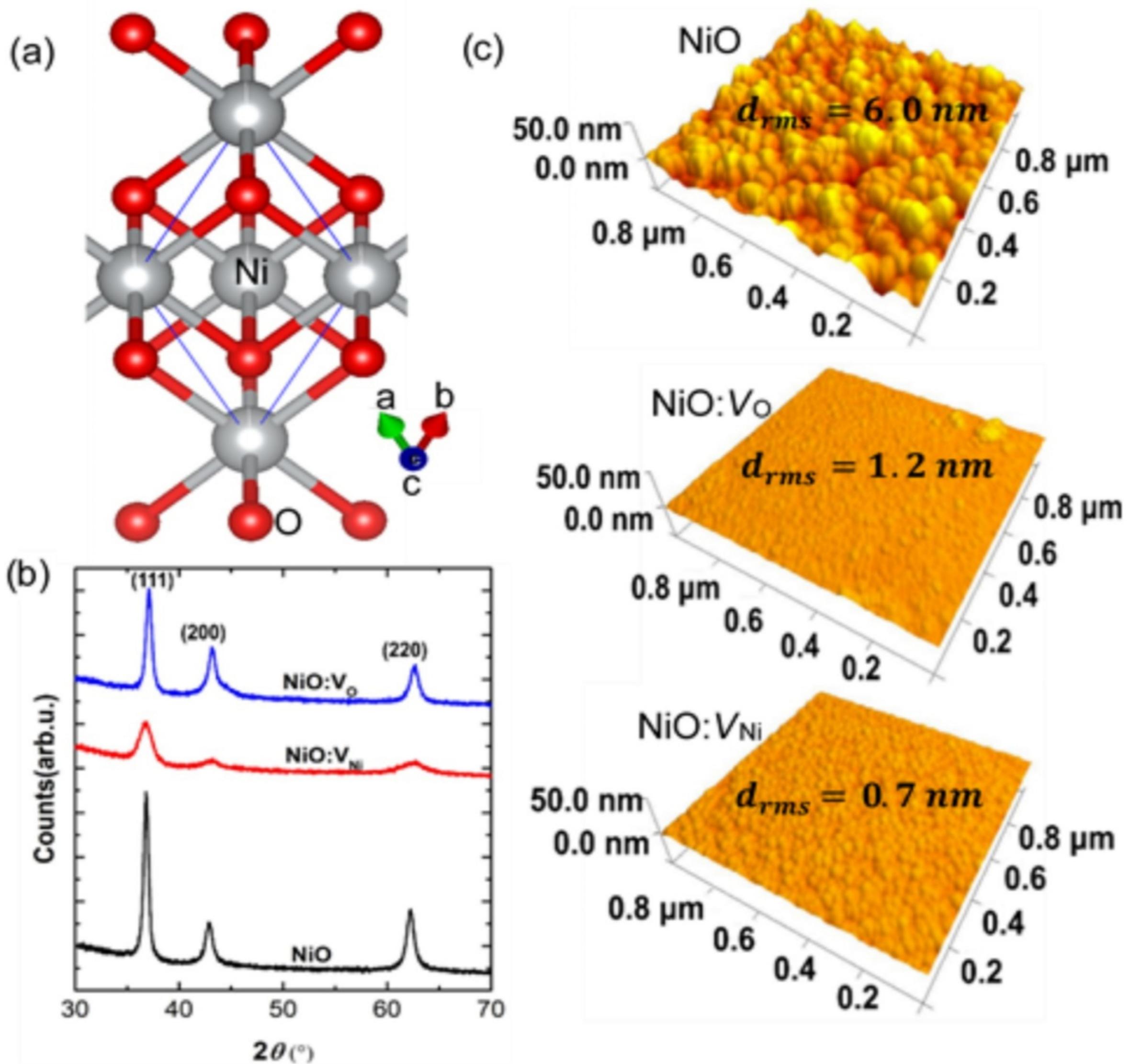

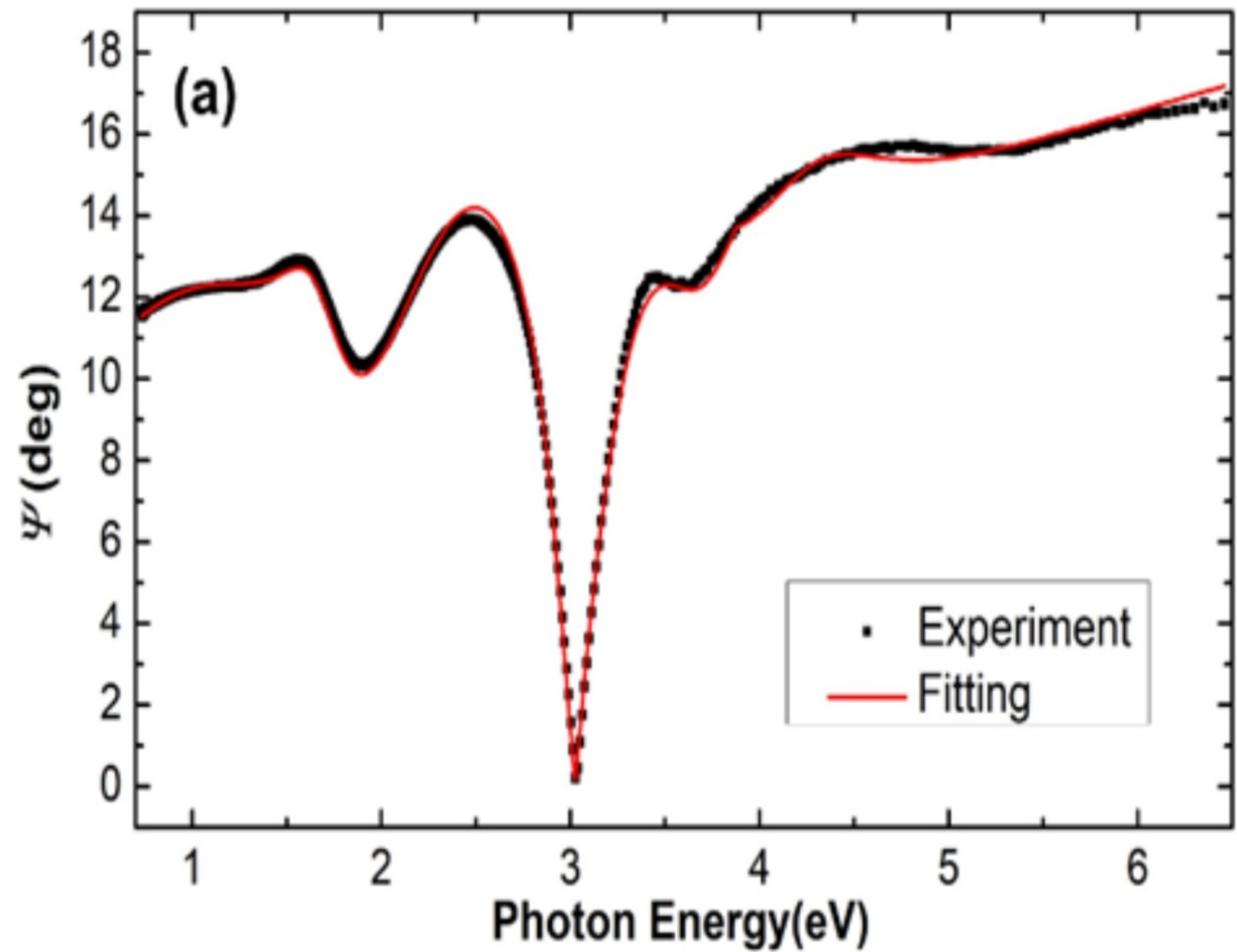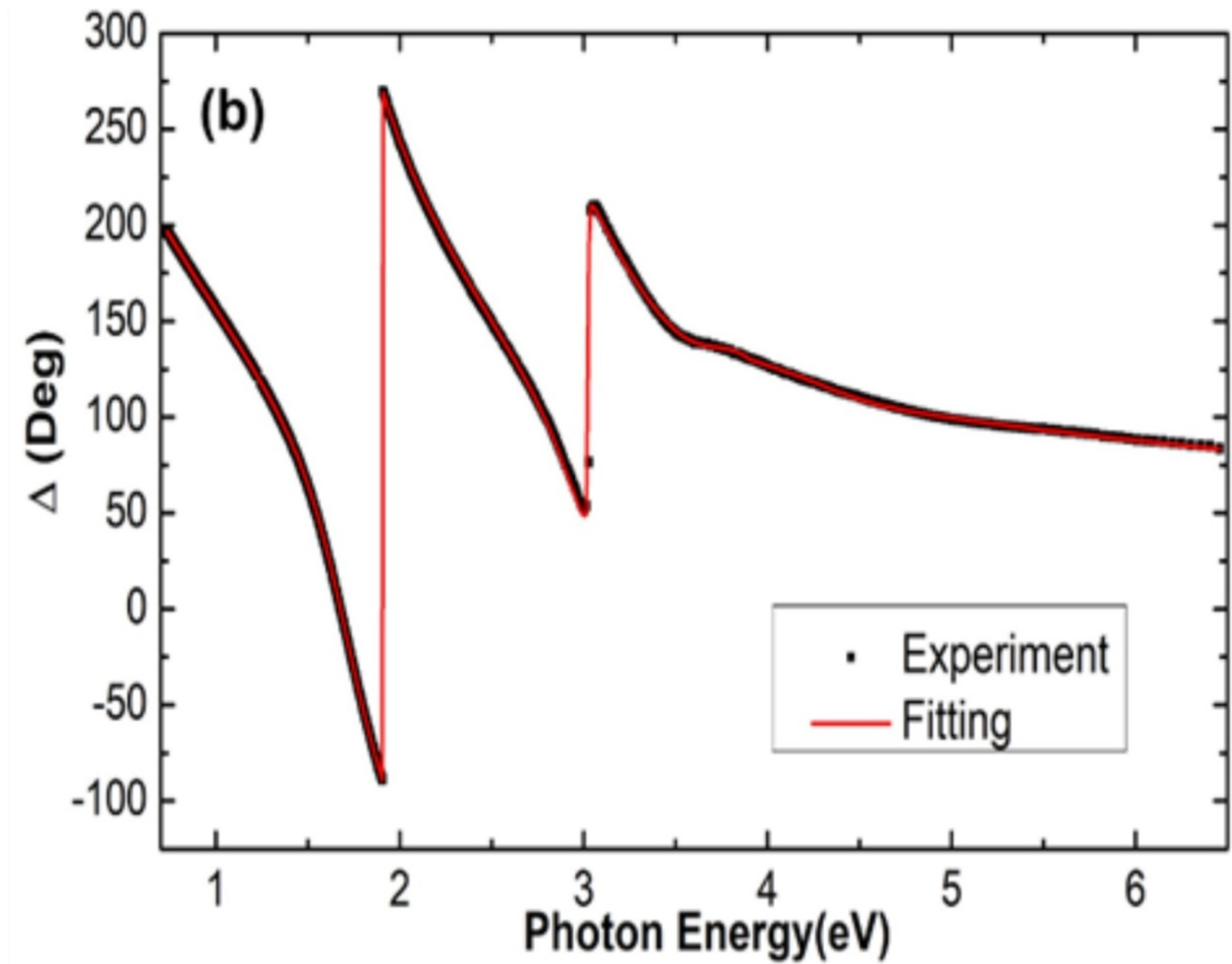

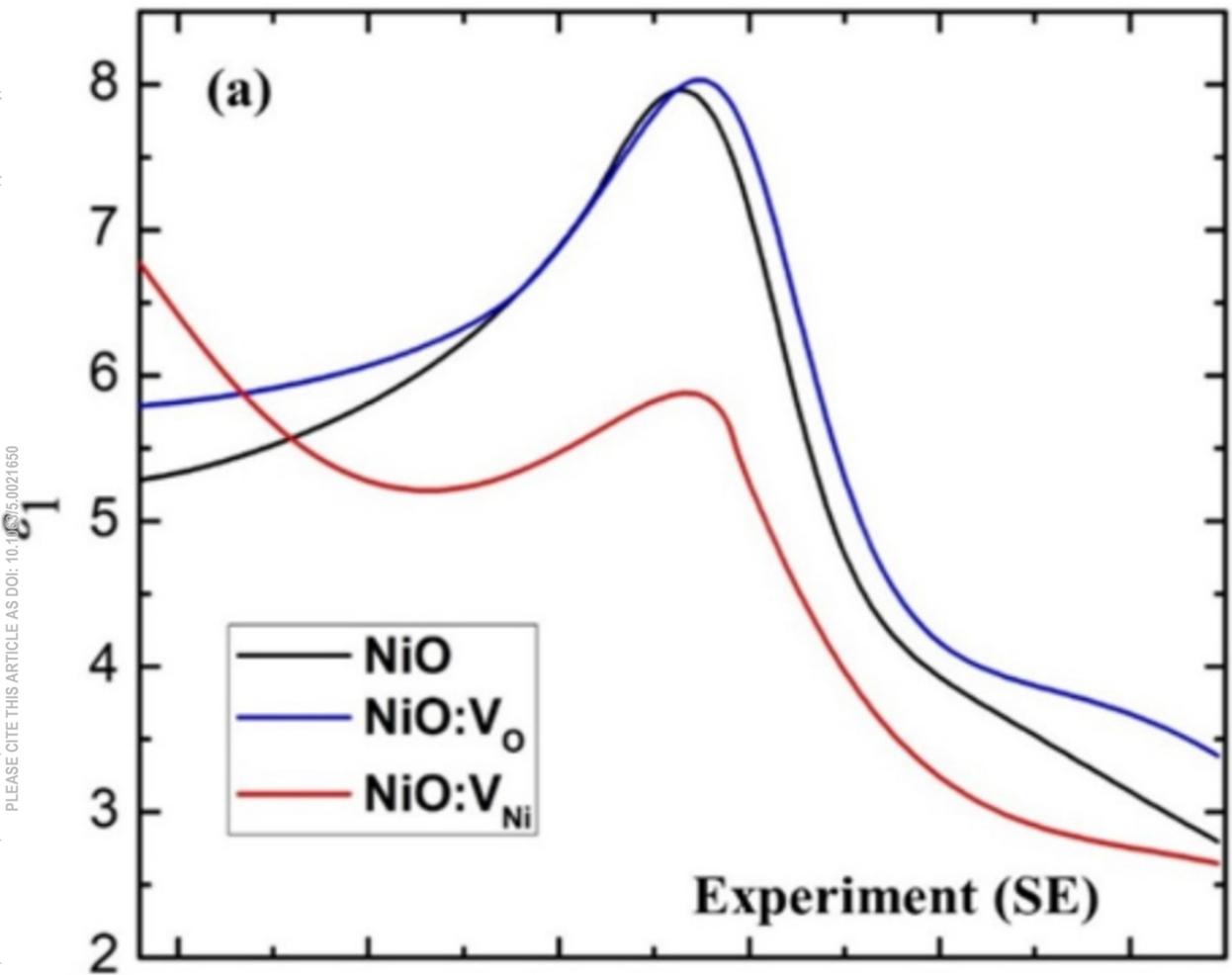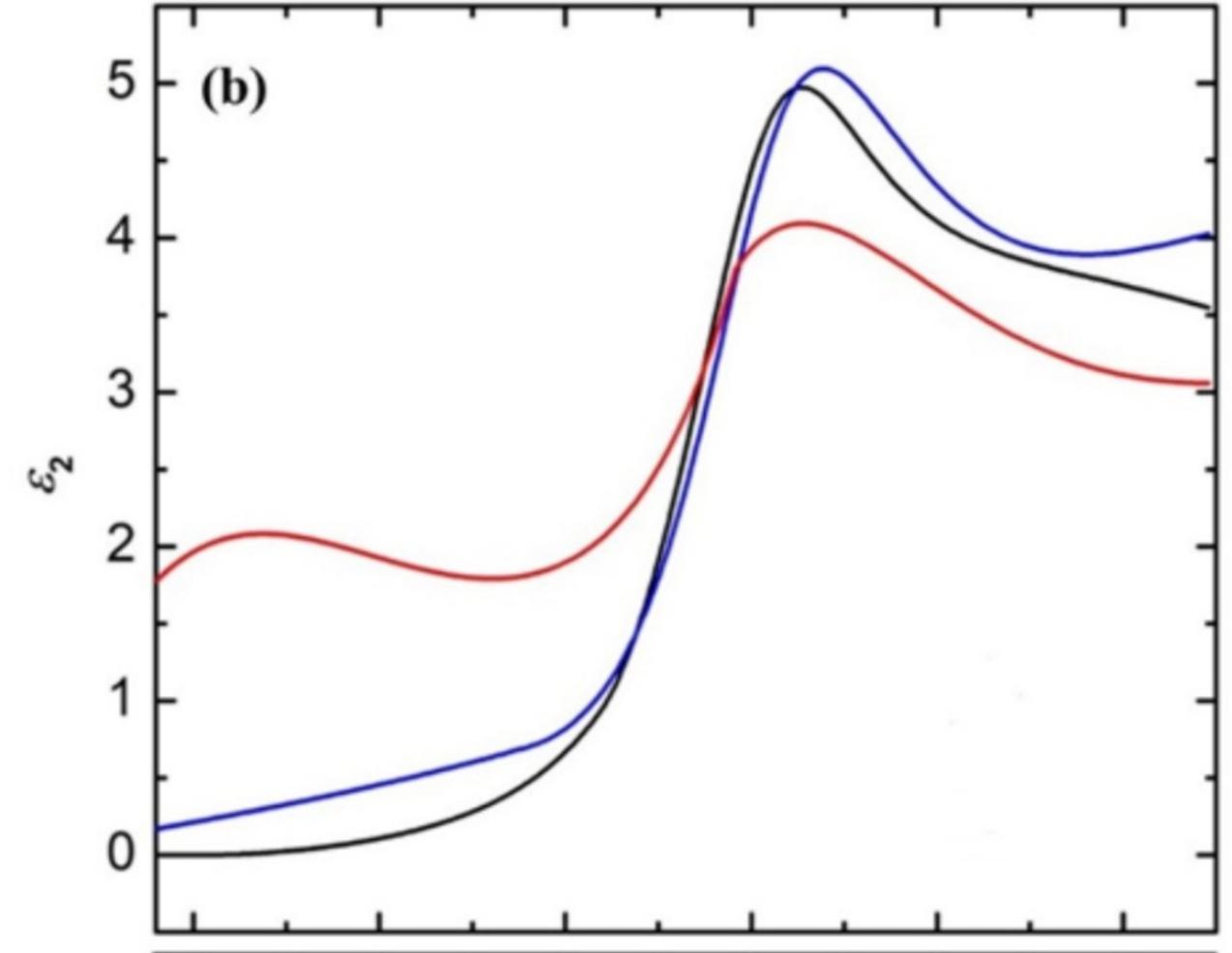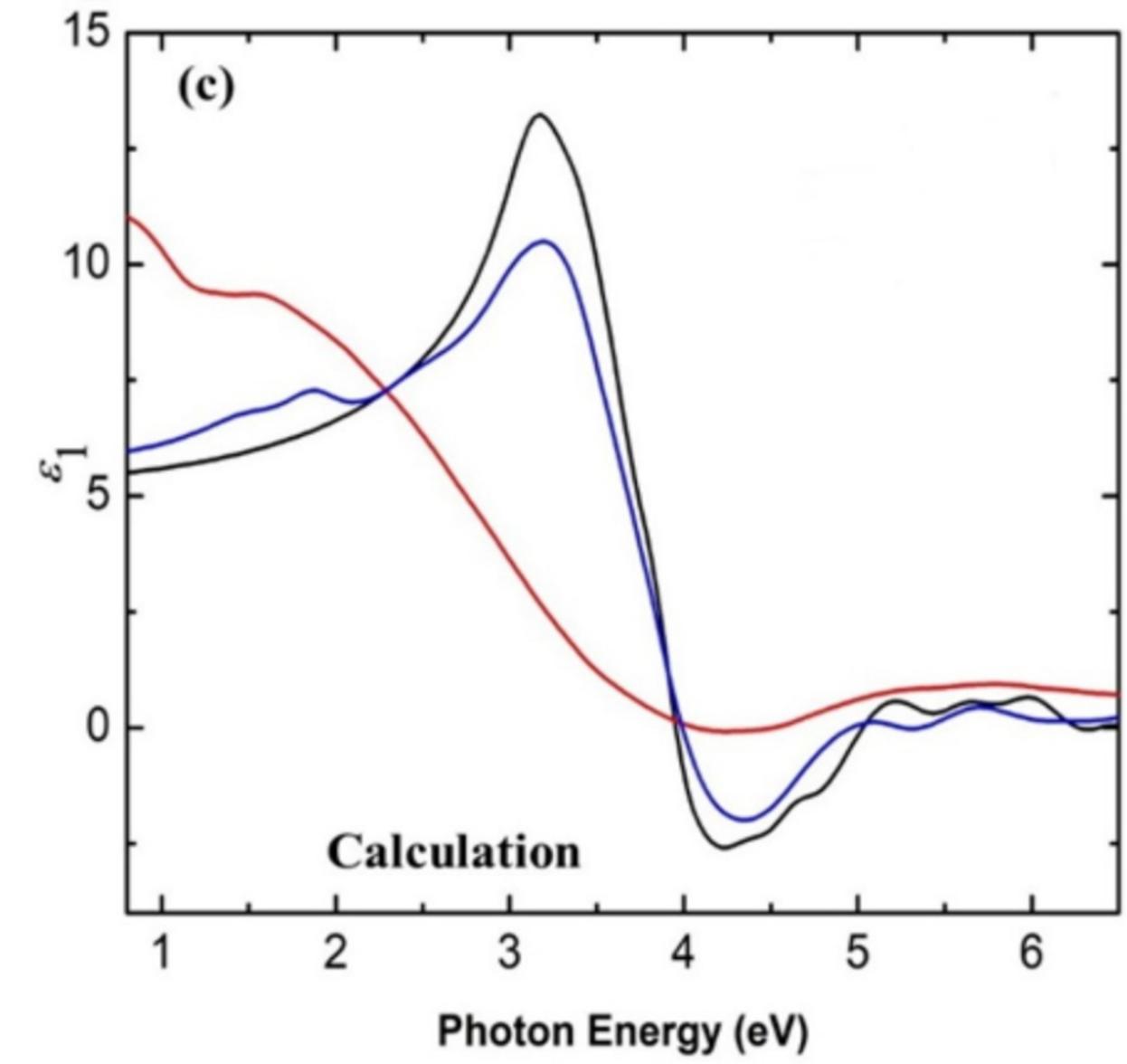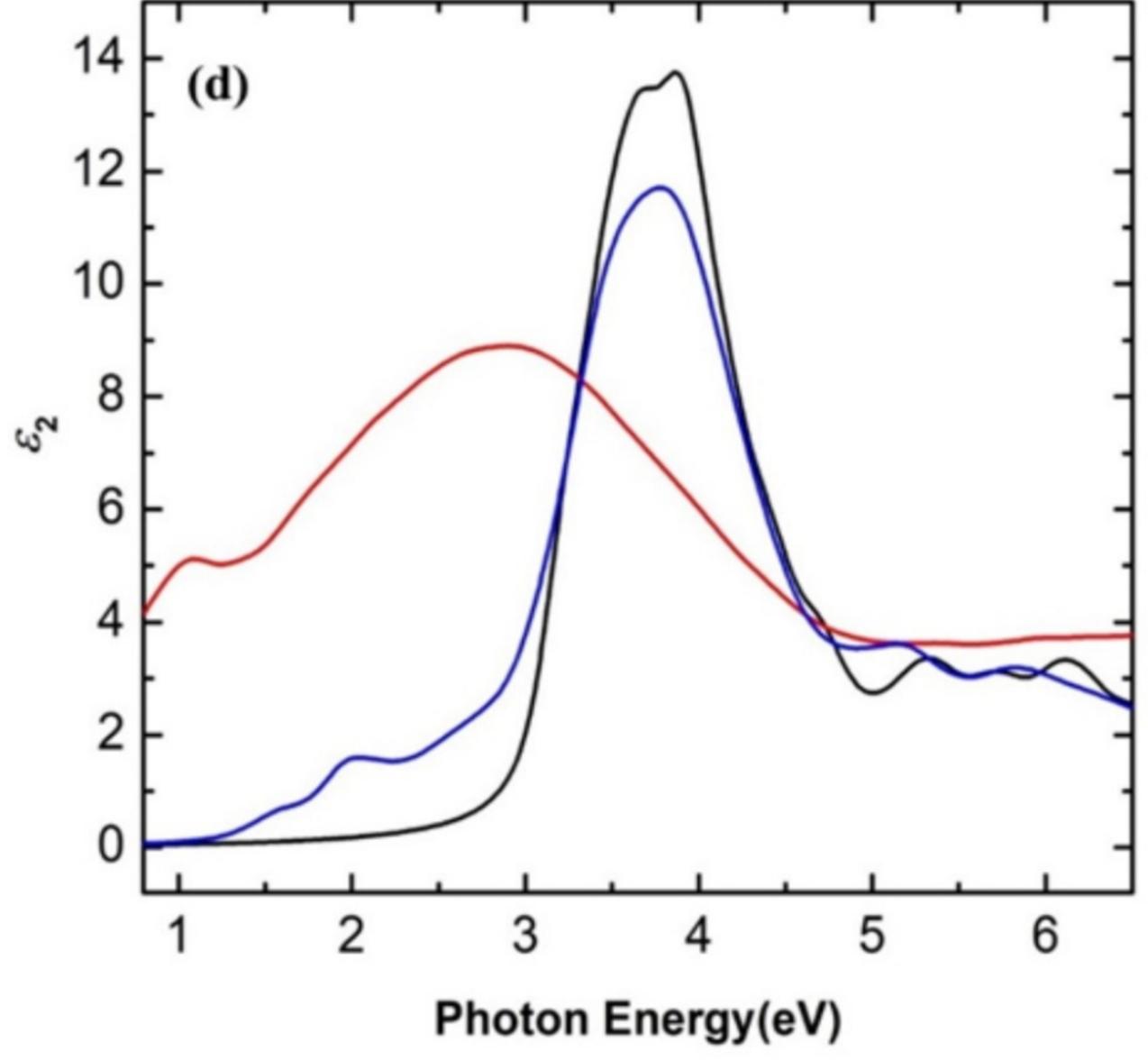

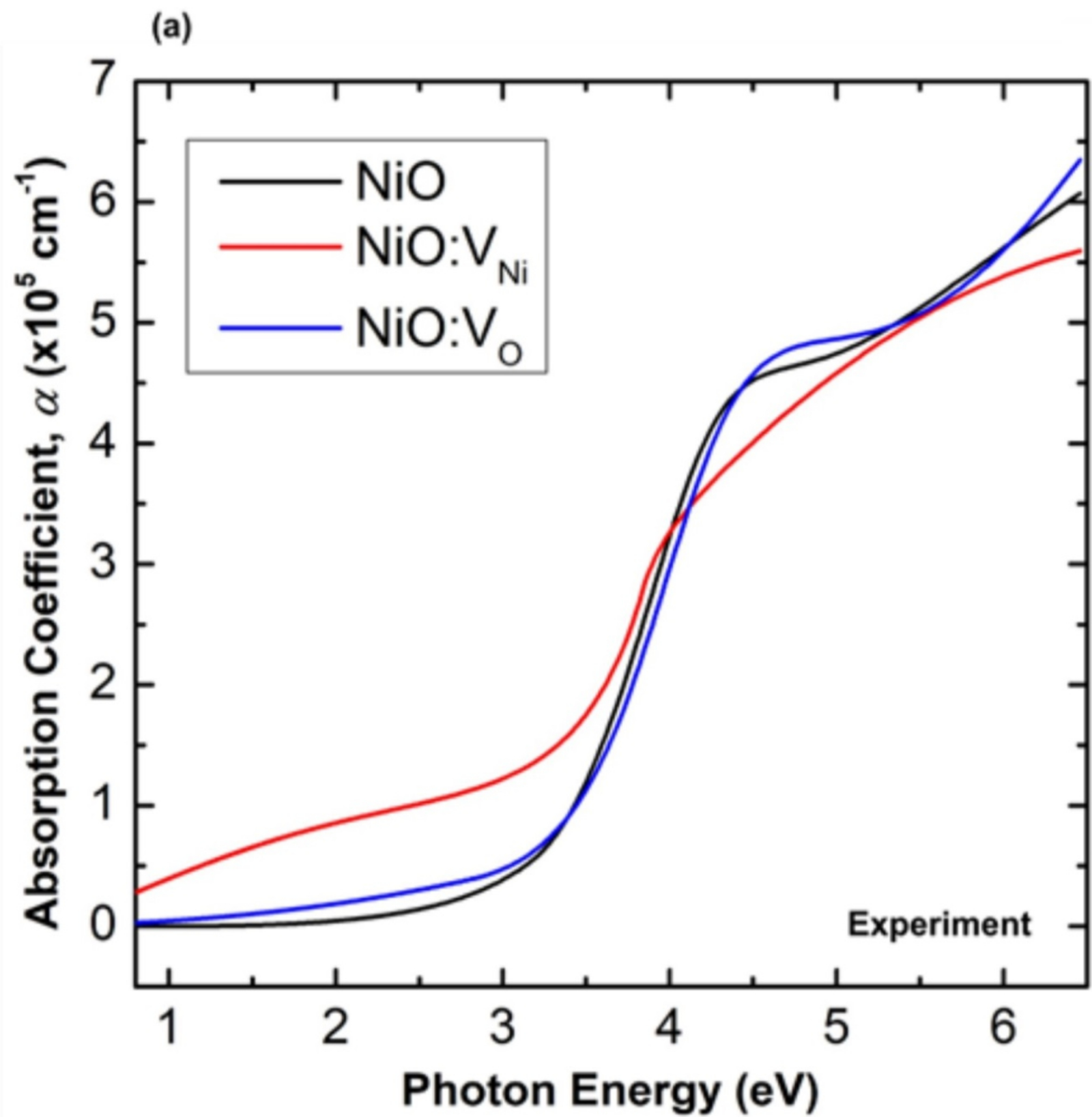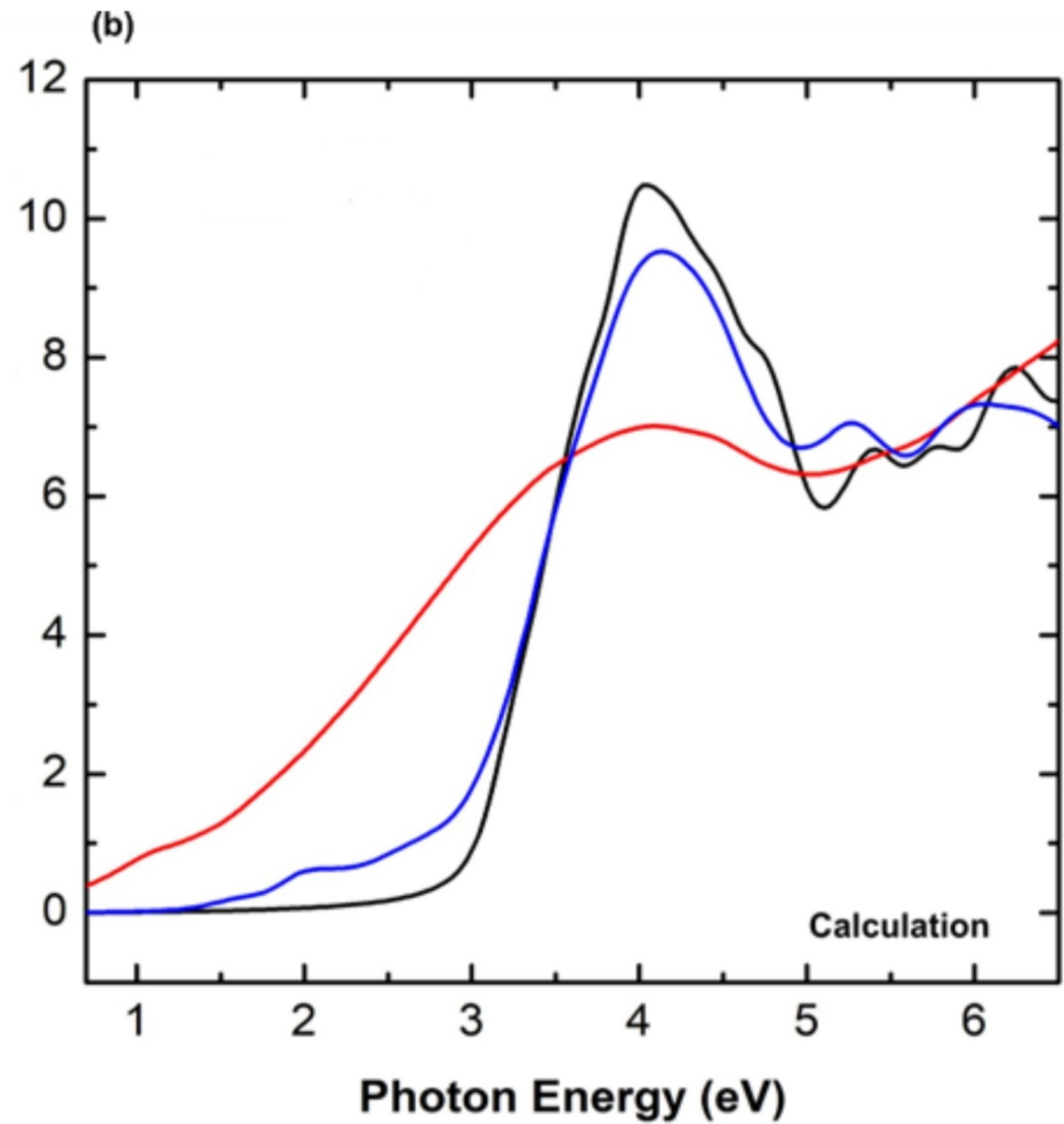

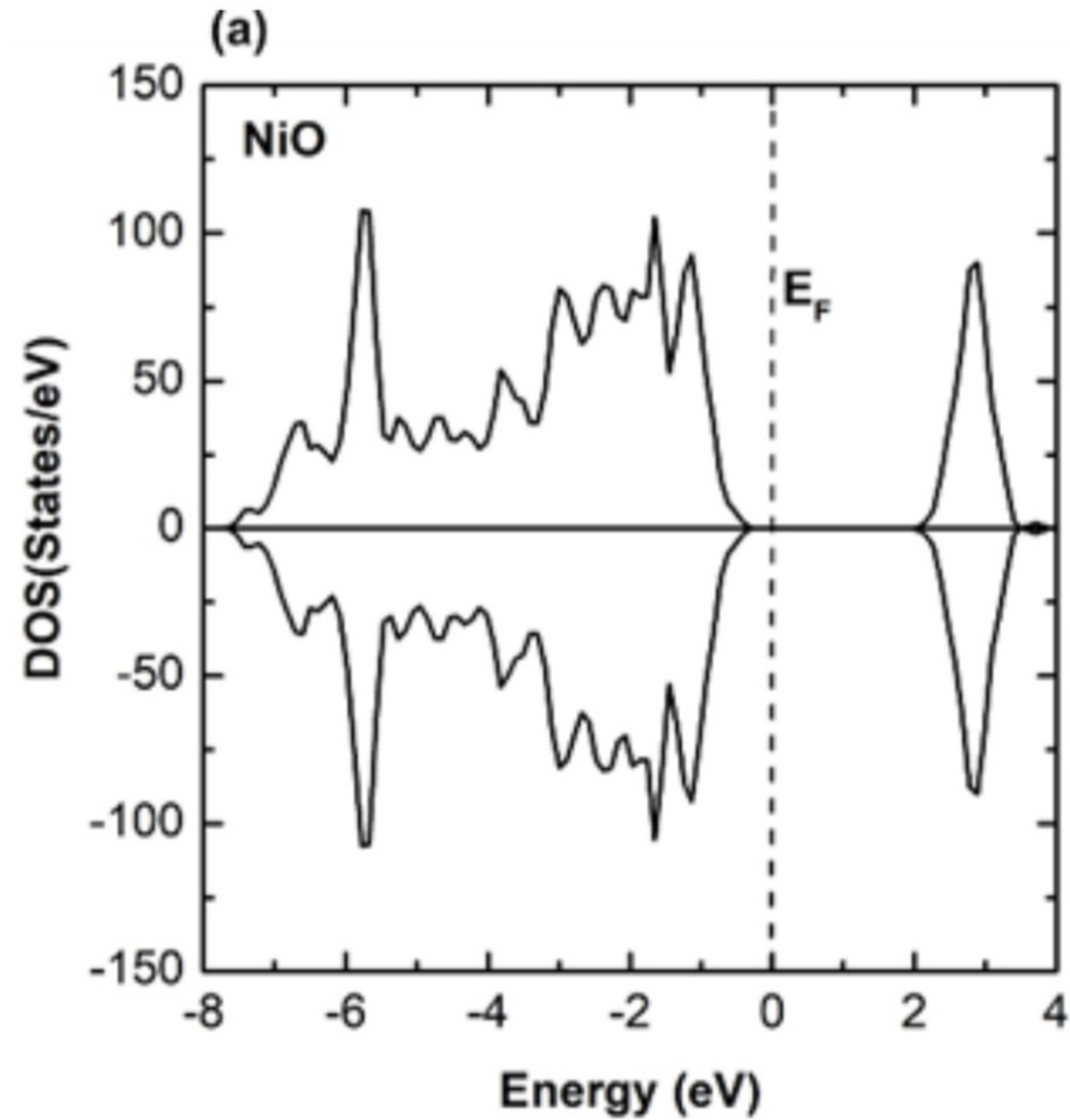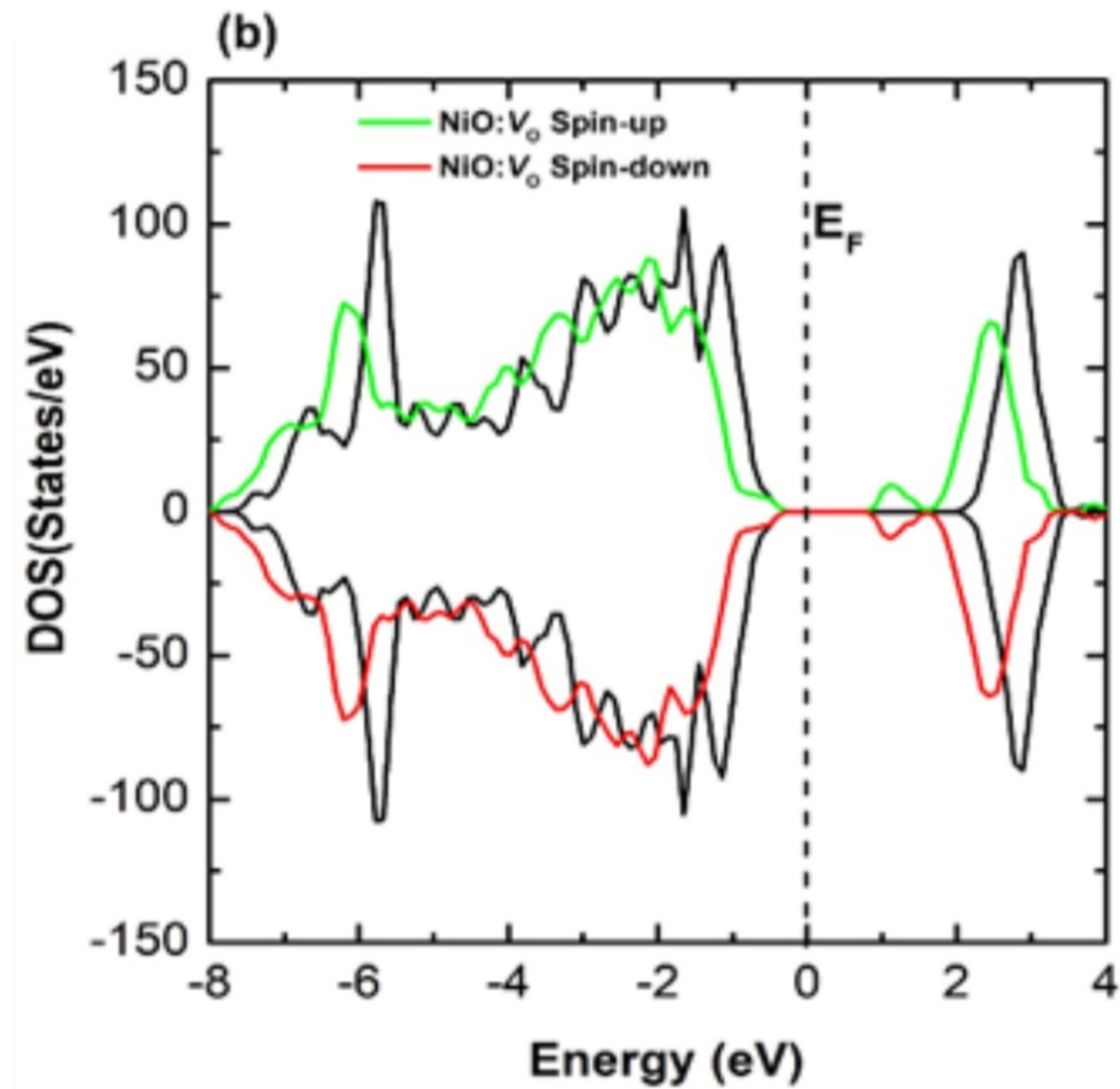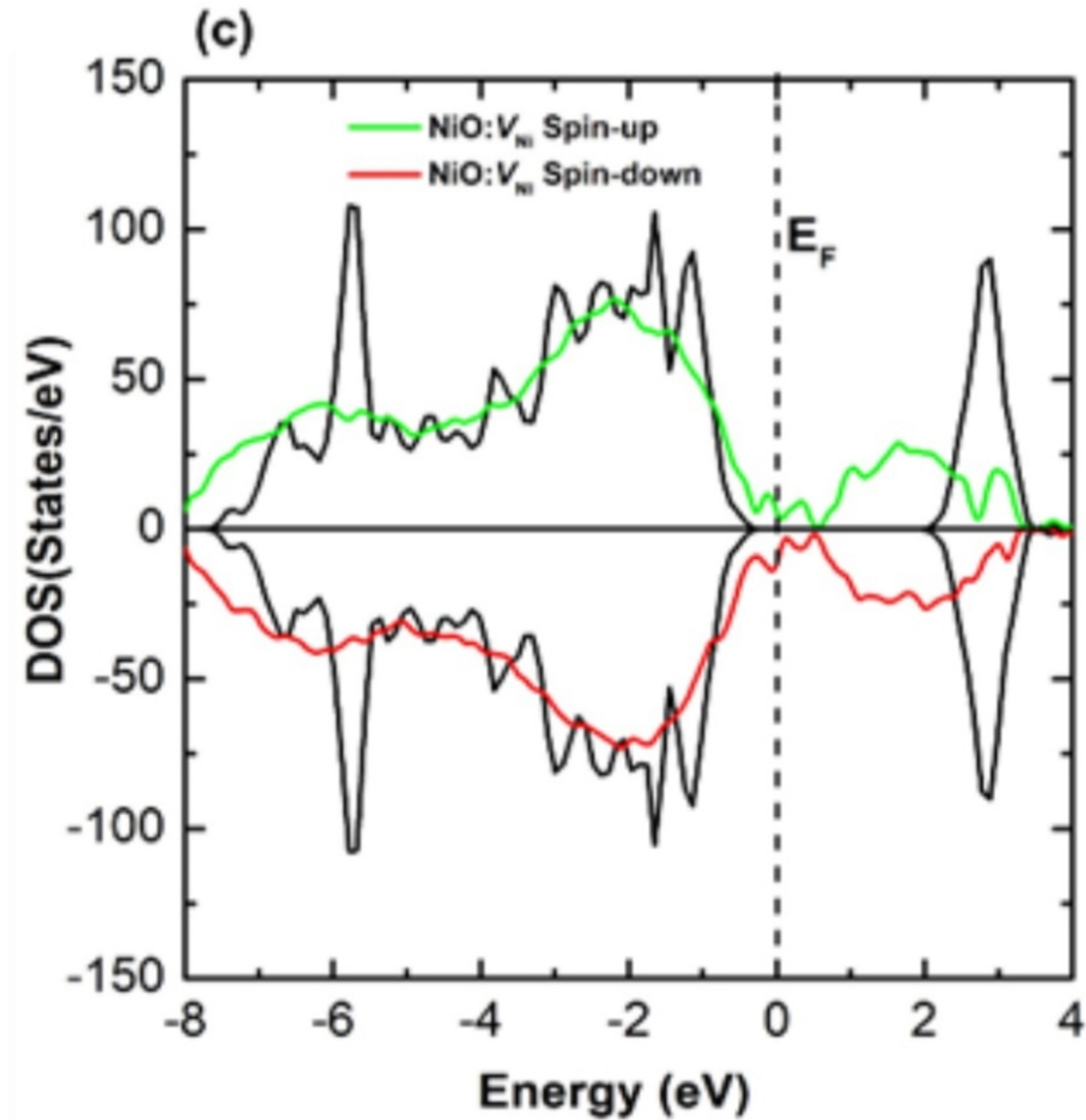

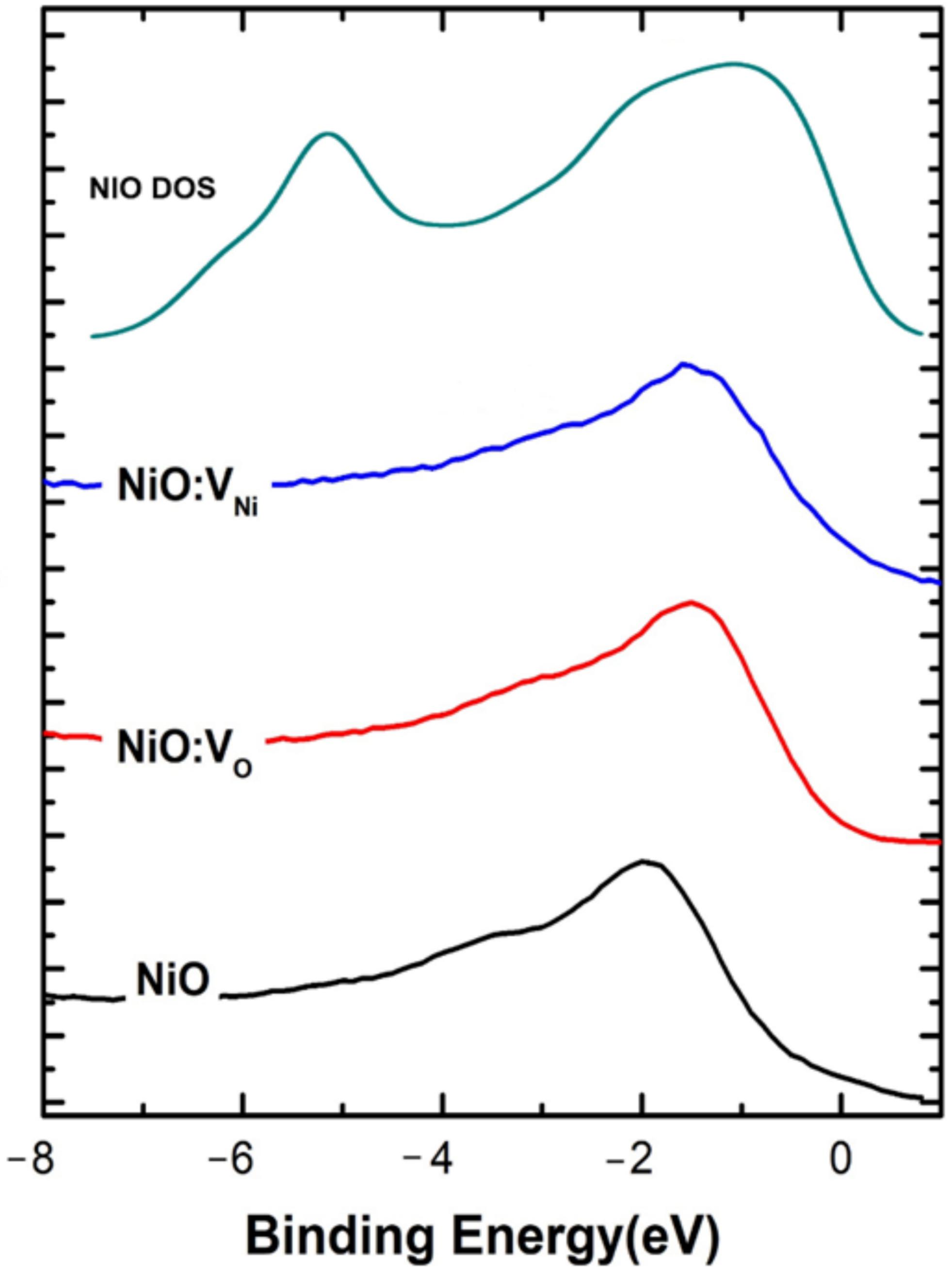